\documentclass[11pt,a4paper]{article} 
\usepackage[utf8]{inputenc} 
\usepackage[T1]{fontenc}
\usepackage{amsmath,amsfonts,latexsym,amssymb,hhline,stmaryrd,color,verbatim,graphicx,epstopdf,slashed,multirow}
\usepackage[mode=text]{siunitx}
\usepackage{jheppub}
\usepackage{url}
\pdfoutput=1

\newcommand{\citere}[1]{{Ref.~\cite{#1}}}
\newcommand{\FR}{{\sc FeynRules}}
\newcommand{\MG}{{\sc MG5\_aMC}}
\newcommand{\MGfull}{{\sc MadGraph5\_aMC@NLO}}
\newcommand{\PY}{{\sc Pythia}}
\newcommand{\DEL}{{\sc Delphes}}
\newcommand{\FJ}{{\sc FastJet}}
\newcommand{\MET}{{\slashed{E}_T}}
\newcommand{\ie} {{\it i.e.\ }}
\newcommand{\eg} {{\it e.g.\ }}
\newcommand*{\Lag}{\ensuremath{\mathcal{L}}}
\newcommand{\lamchi}{\lambda_\text{HP}}
\newcommand{\lamX}{\lambda_\text{HP}}
\newcommand{\lamB}{\lambda_\text{HP}}

\newcommand{\sigmaIND}{\sigma_{\mathrm{B}}^{\mathrm{ind}}} 
\newcommand{\sigmaSYS}{\sigma_{\mathrm{B}}^{\mathrm{sys}}} 
\newcommand{\D}{\mathrm{d}}
\newcommand{\I}{\mathrm{i}}
\newcommand{\tot}{\mathrm{tot}}

\newcommand{\refeq}[1]{{Eq.\;(\ref{#1})}}

\preprint{CP3-19-62, TTK-19-54, MITP/19-086} 

\title{Probing Higgs-portal dark matter with vector-boson fusion}

\author[a]{Jan Heisig}
\author[b]{\!\!, Michael Kr\"{a}mer}
\author[c]{\!\!, Eric Madge}
\author[b]{\!\!, Alexander M\"uck}

\affiliation[a]{Centre for Cosmology, Particle Physics and Phenomenology (CP3),\\ Universit\'e catholique de Louvain, Chemin du Cyclotron 2, B-1348 Louvain-la-Neuve, Belgium}
\affiliation[b]{Institute for Theoretical Particle Physics and Cosmology, \\ RWTH Aachen University, D-52056 Aachen, Germany}
\affiliation[c]{PRISMA$^+$  Cluster of Excellence and Mainz Institute for Theoretical Physics, \\ Johannes  Gutenberg-Universit\"at  Mainz, 55099 Mainz, Germany}

\emailAdd{jan.heisig@uclouvain.be}
\emailAdd{mkraemer@physik.rwth-aachen.de}
\emailAdd{eric.madge@uni-mainz.de}
\emailAdd{mueck@physik.rwth-aachen.de}

\abstract{
We constrain the Higgs-portal model employing the vector-boson fusion channel at the LHC\@.
In particular, we include the phenomenologically interesting parameter region near the Higgs resonance, where 
the Higgs-boson mass is close to the threshold for dark-matter production and a running-width
prescription has to be employed for the Higgs-boson propagator. Limits for the Higgs-portal coupling
as a function of the dark-matter mass are derived from the CMS search for invisible Higgs-boson 
decays in vector-boson fusion at \SI{13}{\TeV}.
Furthermore, we perform projections for the \SI{14}{\TeV} HL-LHC and the \SI{27}{\TeV} HE-LHC 
taking into account a realistic estimate of the systematic uncertainties. 
The respective upper limits on the invisible branching ratio of the Higgs boson reach a level of 
\SI{2}{\percent} and constrain perturbative Higgs-portal couplings up to dark-matter masses of about \SI{110}{\GeV}.
}

\begin{document}
\maketitle


\section{Introduction}\label{sec:intro}

The existence of dark matter (DM) constitutes one of the main puzzles in modern physics. It has stimulated 
a broad range of experimental tests, ranging from direct and indirect detection experiments 
to searches for DM candidates at the LHC~\cite{Bertone:2010zza}.

Minimal Higgs-portal models are particularly simple,
supplementing the Standard Model (SM) by the DM field only~\cite{Silveira:1985rk,McDonald:1993ex,Burgess:2000yq}. The interaction of DM with the SM is mediated by the SM Higgs boson alone. 
We concentrate on the case of scalar DM, which has the compelling feature that the portal interaction
is renormalizable. For higher-spin choices of the DM field, the model can be considered as an effective description, necessitating the 
introduction of further fields to restore renormalizability and unitarity at high energies~\cite{Kim:2008pp,LopezHonorez:2012kv,Baek:2012se,Walker:2013hka,Freitas:2015hsa}.

The model is phenomenologically attractive if the DM mass satisfies $m_\text{DM}\sim m_h/2$ such that DM annihilation is 
resonantly enhanced. This region of parameter space is of particular interest as it 
reconciles the relic density constraint\footnote{See \citere{Binder:2017rgn,Ala-Mattinen:2019mpa} for an improved calculation of the
relic abundance with focus on the Higgs-boson resonance.} and the strong limits from direct detection.
As a matter of fact, the resonance region constitutes one of the two preferred regions in global fits of the model 
(see \eg \cite{Athron:2017kgt,Athron:2018ipf}), 
taking into account constraints from the relic density, invisible Higgs decays, direct and indirect detection.
While current limits from direct detection~\cite{Aprile:2018dbl} exclude the region above the resonance up to $m_\text{DM}\sim \SI{1}{\TeV}$, indirect detection potentially imposes relevant constraints only in a narrow window~\cite{Feng:2014vea},
as the velocity-averaged annihilation rate today peaks sharply around
$m_\text{DM}=m_h/2\,\pm\, \mathcal{O}(\Gamma_\text{tot})$, where $\Gamma_\text{tot}$ is the total Higgs width.
However, parts of this region
turn out to be preferred when fitting an indirect detection signal within the model, such as
the gamma-ray Galactic center excess or the cosmic-ray antiproton excess~\cite{Cuoco:2016jqt,Cuoco:2017rxb}. In this case the fit is sensitive
to changes of the DM mass of the order of $\Gamma_\text{tot}$.\footnote{%
Similar results are found in global fits within other models with Higgs mediated interactions~\cite{Eiteneuer:2017hoh,Arina:2019tib}.}

While LHC searches are not sensitive to couplings that lead to the measured relic density in a canonical freeze-out scenario for $m_\text{DM}\gtrsim m_h/2$, larger couplings might be realized in nature, when deviating from the standard scenario. For instance, the scalar particle may not make up all of DM\@. This case could even be preferred for DM masses around the resonance as discussed in~\cite{Cuoco:2016jqt,Cuoco:2017rxb}.
In alternative models, the scalar particle is not itself the DM candidate but a co-annihilating partner within a dark sector. In this case the region of very efficient annihilation (that would lead 
to highly under-abundant DM in the canonical scenario) is of particular interest. It opens up the possibility of achieving the measured relic density for example via conversion-driven freeze-out~\cite{Garny:2017rxs,DAgnolo:2017dbv}. In this case the metastable singlet scalar would escape the detector invisibly. A similar scenario has been considered in~\cite{Maity:2019hre}.
Deviating even further from canonical assumptions, the coupling required by the relic density constraint can be largely altered by a non-standard cosmological history~\cite{Kamionkowski:1990ni}, while direct detection limits can be relaxed in minimal extensions of the singlet scalar Higgs portal model~\cite{Gross:2017dan,Casas:2017jjg,Bhattacharya:2017fid}. This potentially reopens the parameter space above the resonance and renders collider experiments to be a unique probe of the model.	

In this work, we derive limits on the Higgs-portal coupling from an LHC search in the 
vector-boson fusion (VBF) channel~\cite{Jones:1979bq,Cahn:1983ip} which is a particularly promising channel to search for DM with couplings to 
the SM Higgs boson only~\cite{Eboli:2000ze}. Several Higgs production channels have been investigated in the past~\cite{Craig:2014lda,Endo:2014cca,Han:2016gyy,Dercks:2018wch,Arcadi:2019lka} 
and it has been shown that the VBF channel is the most sensitive one, motivating extensive studies of this channel 
in searches for invisible Higgs decays and Higgs-portal DM at current and future 
colliders~\cite{Bernaciak:2014pna,Goncalves:2017gzy,Biekotter:2017gyu,Buttazzo:2018qqp,Sirunyan:2018owy,Aaboud:2018sfi,Ruhdorfer:2019utl}.

As we have argued before, the Higgs resonance is of particular interest. 
However, LHC studies have either considered the on-shell regime, constrained by the invisible Higgs 
branching ratio (see \eg \citere{Sirunyan:2018owy,Aaboud:2018sfi} for recent experimental results), 
or heavier DM particles which are produced via a
highly off-shell Higgs boson~\cite{Craig:2014lda,Endo:2014cca,Ruhdorfer:2019utl}.  
In this paper we close this gap by calculating limits on the Higgs-portal model with a special emphasis on 
analysing the region with $m_\text{DM}\sim m_h/2$. For $m_\text{DM} = m_h/2 \pm \mathcal{O}(\Gamma_\text{tot})$
and sizeable DM couplings, the total Higgs width as a function of the invariant mass  
varies substantially and distorts the Higgs-boson line-shape due to the opening of the Higgs decay into DM\@. As a consequence,
a fixed-width computation becomes unreliable and needs to be improved using a running width 
in the Higgs propagator.

We reinterpret the \SI{13}{\TeV} VBF analysis for invisble Higgs-boson decays 
by CMS~\cite{Sirunyan:2018owy} to establish limits on the Higgs-portal coupling as a function of the DM mass. In particular, we utilize the bounds 
on the signal strength for the production of an additional invisibly-decaying SM-like Higgs-boson $\cal H$ with mass $m_{\cal H}$ that does not mix with the SM Higgs boson to compute the limits. 
In addition to the reinterpretation of the \SI{13}{\TeV} analysis of~\citere{Sirunyan:2018owy}, we derive prospects for the \SI{14}{\TeV} HL-LHC 
on the basis of~\citere{CMS:2018tip}, and for a possible \SI{27}{\TeV} HE-LHC upgrade. In contrast to
considering ultimate sensitivities~\cite{Craig:2014lda,Endo:2014cca,Ruhdorfer:2019utl},
we put particular emphasis on estimating the systematic uncertainties on the data-driven background prediction.
As a by-product of our analysis within the  singlet scalar Higgs-portal model, in analogy
to~\citere{Sirunyan:2018owy}, we derive projected limits on the on-shell production of an additional Higgs-boson $\cal H$. 
These results might be useful to constrain other models using the procedure outlined in this work.

The remainder of this paper is organized as follows. 
In Sec.~\ref{sec:model} we briefly introduce the singlet scalar model. 
In Sec.~\ref{sec:13TeVlim} we derive current \SI{13}{\TeV} constraints.
Projections for the HL- and HE-LHC are studied in Sec.~\ref{sec:projections}.
We conclude in Sec.~\ref{sec:summary}. 
Appendix~\ref{app:res} provides more details on the running-width prescription 
and Appendix~\ref{app:othmod} discusses alternative choices regarding the spin of the DM candidate.

\section{Higgs portal model}\label{sec:model}

We consider the scalar singlet Higgs-portal model~\cite{Silveira:1985rk,McDonald:1993ex,Burgess:2000yq}, which is among the simplest possible UV-complete extensions of the SM\@.
It extends the SM by a real singlet scalar field $S$ that is stabilized by a $Z_2$ symmetry and thereby provides a DM candidate.
The corresponding Lagrangian is 
\begin{equation}
\label{eq:Lagrangian}
	\mathcal{L} = \mathcal{L}_\text{SM} + \frac{1}{2} \partial_\mu S\, \partial^\mu S - \frac{1}{2} m_{S,0}^2 S^2 - \frac{1}{4} \lambda_S^2 S^4 - \frac{1}{2} \lambda_\text{HP} S^2\,\Phi^\dagger \Phi \, ,
\end{equation}
where $\Phi$ is the SM Higgs doublet.
After electroweak symmetry breaking, in unitary gauge we can write $\Phi = (0,h+v)/\sqrt{2}$,
where $v\simeq\SI{246}{\GeV}$ is the SM Higgs vacuum expectation value and the scalar mass is given by $m_S^2 = m_{S,0}^2 + \lambda_\text{HP} v^2/2$. The Higgs-portal coupling induces the interactions
\begin{equation}
{\cal L}_\text{HP} = - \frac{1}{4} \lambda_\text{HP}\, h^2 S^2 - \frac {1}{2} \lambda_\text{HP} v\, h\, S^2\,,
\label{eq:ewbr}
\end{equation}
where the latter is relevant for the phenomenology considered here.

If $m_S < m_h/2$, the Higgs boson can decay invisibly into two DM scalars with the decay width
\begin{equation}
\label{eq:width}
	\Gamma_\text{inv} = \frac{\lambda_\text{HP}^2 v^2}{32 \pi m_h} \sqrt{1-4\frac{m_S^2}{m_h^2}}\,.
\end{equation}

For an extensive discussion of the DM and collider phenomenology of this and other Higgs-portal models,
we refer the reader to the recent and comprehensive review \citere{Arcadi:2019lka} and references therein.

\section{LHC limits at 13 TeV}\label{sec:13TeVlim}

\begin{figure}[t]
\centering\includegraphics[bb=0 0 160 100]{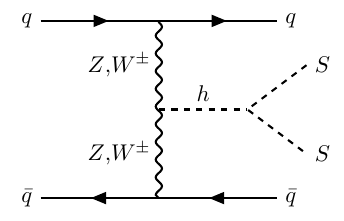}
\caption{
Feynman diagram for Higgs-portal DM production in VBF at the LHC.
\label{fig:feyn}	
}
\end{figure}

In this section, we derive limits on the Higgs-portal coupling $\lambda_\text{HP}$ in \refeq{eq:Lagrangian} using
the results of the \SI{13}{\TeV} CMS search for invisible Higgs decays in the VBF channel as reported in~\citere{Sirunyan:2018owy}. 
The corresponding ATLAS study can be found in~\citere{Aaboud:2018sfi}.
In VBF, the Higgs boson is produced in association with two jets (see Figure~\ref{fig:feyn}) that are characterised by a large separation in pseudorapidity and by a large dijet invariant 
mass. 
The search presented in \citere{Sirunyan:2018owy} is based on a cut-and-count and on 
a shape analysis, where the shape of the dijet invariant mass is used to impose limits on the invisible branching 
ratio of the SM Higgs boson. In addition, limits on the production cross section for a
SM-like heavy Higgs boson are reported, assuming a branching ratio of one for the decay
into invisible final states and no mixing with the SM Higgs.

In Section~\ref{sec:reinterp}, we derive limits on the Higgs-portal coupling $\lambda_\text{HP}$ exploiting all the 
results of~\citere{Sirunyan:2018owy}. In particular, we show how the cross-section limits 
on an additional heavy Higgs boson can be used to derive limits on $\lambda_\text{HP}$ for DM masses in the vicinity 
and beyond the threshold, where the on-shell Higgs decay into DM particles is 
kinematically inaccessible. We employ both the cut-and-count analysis as well as the shape analysis. 

In Section~\ref{sec:recast}, we recast the cut-and-count analysis of~\cite{Sirunyan:2018owy}
using leading-order~(LO) Monte Carlo simulations. Employing a simple rescaling of the LO cross sections to match the 
more sophisticated predictions in~\citere{Sirunyan:2018owy}, we reproduce the bounds on the 
Higgs-portal coupling as found in Section~\ref{sec:reinterp}. This validation of our Monte Carlo
setup enables us to perform projections for searches at the \SI{14}{\TeV} high-luminosity LHC as well as 
at a \SI{27}{\TeV} high-energy LHC option in Section~\ref{sec:projections}.

\subsection{Reinterpretation of upper limits}\label{sec:reinterp}

In the following, we assume a factorization of Higgs production and decay, \ie we do not consider electroweak corrections or higher-order corrections 
in the Higgs-portal coupling. In this approximation, one can write any fiducial cross
section $\sigma_\text{inv}$ for the production of a pair of DM particles (or any other pair of invisible particles coupling exclusively to the Higgs boson) as  
\begin{equation}
\label{eq:DMproduction}
\sigma_\text{inv} = \int \frac{\D q^2}{2\pi} \,  \sigma_{h}(q^2) \, 
 |P(q^2)|^2 \, 2q\, \Gamma_\text{inv}(q^2) \,\Theta(q^2 - 4m_S^2)\,,
\end{equation}
where $\sigma_{h}(q^2)$ is the corresponding fiducial, detector-level production cross-section 
(including acceptance times efficiency) of the (off-shell) SM 
Higgs boson at a given invariant mass~$q^2$. The Higgs-boson propagator is denoted by $P(q^2)$
and the partial Higgs-boson width $\Gamma_\text{inv}(q^2)$ generalizes the result of \refeq{eq:width}
by replacing the Higgs mass $m_h$ with $\sqrt{q^2}$. Well below threshold ($m_S < m_h/2$), the
on-shell Higgs-boson decay into invisible DM particles is open. Hence, as long as the
Higgs-portal coupling is not too large, the narrow-width approximation applies and 
the right-hand-side of \refeq{eq:DMproduction} reduces to the product of the on-shell Higgs-production
cross-section and the branching ratio into DM pairs. For this case, \citere{Sirunyan:2018owy}
has already interpreted its bound on the invisible branching ratio of the Higgs boson in terms of 
the Higgs-portal model (see Figure 10 of~\citere{Sirunyan:2018owy}). 

In this work, we also address
DM masses in the threshold region and beyond, where DM production involves an
off-shell Higgs boson. In this context, in order to obtain $\sigma_h(q^2)$ from the experimental analysis,
we employ the interpretation of the VBF+MET search 
in terms of an additional SM-like Higgs boson $\cal H$ that does not mix with the \SI{125}{\GeV} 
Higgs boson and decays to invisible particles.
This interpretation provides a limit $\mu_{\mathcal{H}}^{95\%}$ at the \SI{95}{\percent} confidence 
level\footnote{In the analysis of \citere{Sirunyan:2018owy} and throughout 
this work the CL$_\text{s}$ method~\cite{Junk:1999kv,Read:2002hq} 
is used.} (CL) on the signal strength
$\mu_\mathcal{H}= \sigma\times {\cal B} ({\cal H}\to\text{inv}) /\sigma_\text{SM}$ as a
function of the mass $m_{\cal H}$ of the additional Higgs boson for the cut-and-count as well as
the shape analysis (see Figure 7 in 
\citere{Sirunyan:2018owy}). Hence, any BSM contribution $\sigma_\text{inv}$ 
to the cut-and-count measurement 
has to be smaller than the corresponding limit 
\begin{equation}
\label{eq:limitXS}
\sigma_{\text{inv}}^{95\%}=\mu_{\mathcal{H}}^{95\%}(m_\mathcal{H})\sigma_\text{SM}(m_\mathcal{H})=
\mu_{\mathcal{H}}^{95\%}(q^2)\sigma_h(q^2) \, , 
\end{equation}
where we have used 
$\sigma_h(q^2=m^2_\mathcal{H})=\sigma_\text{SM}(m_\mathcal{H})$, \ie the on-shell production 
cross section for an additional SM-like Higgs boson is identical to the off-shell SM Higgs-boson 
cross section at $q^2=m^2_\mathcal{H}$, if electroweak corrections are ignored. Note that 
$\sigma_{\text{inv}}^{95\%}$ is, of course, independent of $m_\mathcal{H}$ or $q^2$.
Using \refeq{eq:limitXS} to eliminate $\sigma_h(q^2)$ in \refeq{eq:DMproduction},  
and dividing by $\sigma_{\text{inv}}^{95\%}$, one finds
\begin{equation}
\label{eq:findbound}
\frac{\sigma_{\text{inv}}}{\sigma_{\text{inv}}^{95\%}} =  \int \frac{\D q^2}{2\pi} \, \frac{1}{\mu^{95\%}_\mathcal{H}(q^2)} \, 
 |P(q^2)|^2 \, 2q \, \Gamma_\text{inv}(q^2) \,\Theta(q^2 - 4m_S^2) \, .
\end{equation}
Since only $\sigma_{\text{inv}}\le \sigma_{\text{inv}}^{95\%}$ is compatible with the measurement
at \SI{95}{\percent} CL, we
can numerically solve \refeq{eq:findbound} to derive the corresponding bound on the 
Higgs-portal coupling for a given DM mass.

We have not yet specified the Higgs-boson propagator.  
It is common to use a fixed-width prescription for the resonant Higgs propagator, \ie
\begin{equation}
P_\text{f}(q^2) = \frac{\I}{q^2-m_h^2+\I\, m_h \,\Gamma_\tot(m_h^2)} \, ,
\end{equation}
where $\Gamma_\tot(m_h^2)$ is the total width of an on-shell Higgs 
boson. However, as discussed in Appendix~\ref{sec:resonance}, a fixed-width prescription 
does not yield a proper description of the threshold region, in particular the transition between 
on-shell and off-shell production of the DM pairs is not properly described if the Higgs-portal
coupling is sizeable. A running-width propagator
\begin{equation}
\label{eq:runningWidth}
P_\text{r}(q^2) = \frac{\I}{q^2-m_h^2+\I \sqrt{q^2} \,\Gamma_\tot(q^2)} \, ,
\end{equation}
solves this issue (see Appendix~\ref{sec:resonance}) 
and yields an improved description of the threshold region.

\begin{figure}[t]
\centering
\includegraphics[bb=0 0 450 330]{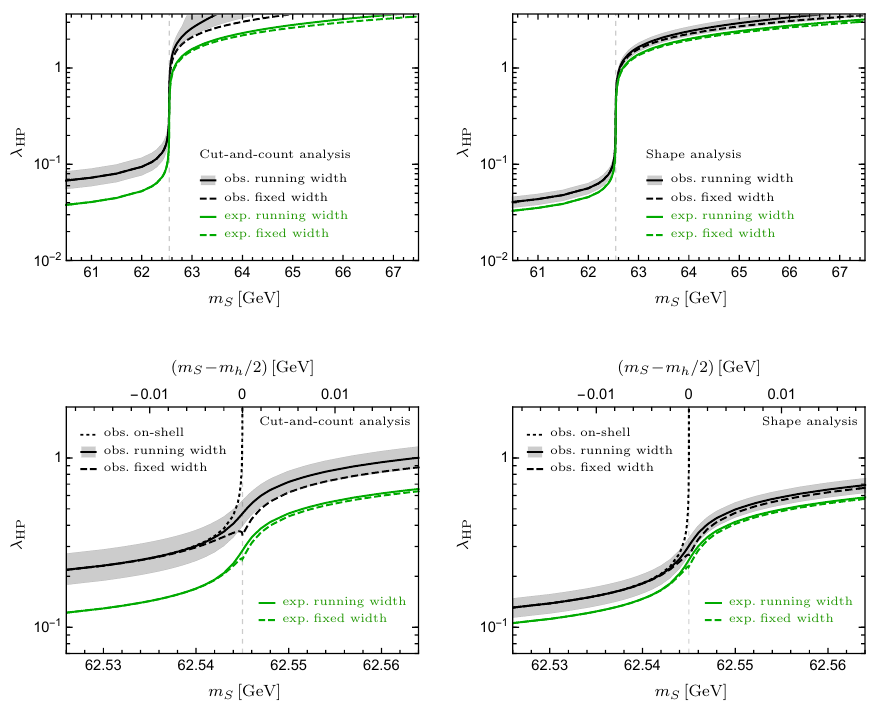}
\caption{%
Upper limits at the \SI{95}{\percent}~CL on the Higgs-portal coupling $\lambda_\text{HP}$ as a function
of the DM mass $m_S$. We present results for the cut-and-count analysis (left) and the shape
analysis (right) of \citere{Sirunyan:2018owy}. We show a wide range for the dark-mater mass (top) as
well as a zoom into the most interesting threshold region (bottom). In all plots, we show expected as
well es observed limits employing the fixed- and the running-width description for the Higgs propagator.
The grey band around the running-width observed limit indicates 
the \SI{17}{\percent} uncertainty on the Higgs production cross-section stated in~\citere{Sirunyan:2018owy}.
In the threshold-region plots (bottom), we also indicate the limits due to the decay of an on-shell
Higgs boson into DM pairs.
\label{fig:limits13TeV}
}
\end{figure}

We use Figure 7 in \citere{Sirunyan:2018owy} to read off $\mu_\mathcal{H}(q^2)$.\footnote{Although
we are confident to extract the data with negligible error, we would highly appreciate to
find data like this on HEPData~\cite{hepdata}.}
Solving~\refeq{eq:findbound} numerically,
we obtain the bounds from the cut-and-count analysis as shown in the left panels of 
Figure~\ref{fig:limits13TeV}. Below threshold, the Higgs-portal coupling is constrained to small values 
below $\sim0.1$ and the limits are identical to the limits obtained from the experimental limit on the 
invisible branching ratio of the SM Higgs boson using \refeq{eq:width} alone. In this region, more
stringent but also more model-dependent bounds from the measurement of the Higgs-boson coupling strength
may be applied to the Higgs-portal model~\cite{Belanger:2013xza,Kraml:2019sis}.  

As expected, the approximation of only employing the invisible branching ratio breaks down at threshold, and our analysis derives the proper limits at and above
threshold. 
At $m_S=m_h/2$ the analysis excludes couplings larger than $\lambda_\text{HP} = 0.47$ at \SI{95}{\percent}~CL\@.
It can also be seen in Figure~\ref{fig:limits13TeV} that the fixed-width description shows an
unphysical feature at threshold.\footnote{Note that we assume $m_h=\SI{125.09}{\GeV}$ throughout this study. However, close to threshold 
the Higgs width is small compared to the experimental error on the Higgs mass and to the characteristic 
energy scales on which  $\sigma_{h}(q^2)$ varies significantly. The decisive quantity upon which the limits depend is thus $m_S-m_h/2$, rather than the absolute value for $m_S$. Therefore we have added the 
respective scale on the upper axes of the lower plots in Figure~\ref{fig:limits13TeV}.}
The difference to the running-width results, as expected, becomes less
prominent for improved limits, \ie a smaller Higgs-portal coupling. Hence, in the expected limits the
effect is less pronounced than in the observed limit. 
Well above threshold this difference, which is formally of higher-order in $\lambda_\text{HP}$, can be viewed as a lower bound on the theoretical 
uncertainty. With increasing DM masses the LHC search rapidly looses sensitivity and can only probe large $\lambda_\text{HP}$ in the non-perturbative regime. 

We also apply \refeq{eq:findbound} to derive limits from the shape analysis. Here, we assume that the
shape of the invariant-mass distribution of the VBF dijet system does not vary substantially with $q^2$.
Note that relevant limits on the Higgs-portal coupling can only be obtained for DM masses for which
the integral in \refeq{eq:findbound} is dominated by values of $q^2$ close to the Higgs mass.
The better sensitivity of the shape analysis translates into improved bounds in the 
Higgs-portal coupling as shown in the right panels of Figure~\ref{fig:limits13TeV}. 
It excludes couplings larger than $\lambda_\text{HP} = 0.30$ for $m_S=m_h/2$.
However, even the shape
analysis cannot constrain perturbative Higgs-portal models for DM masses above \SI{67}{\GeV}.
The effect of the running width is still noticeable in the observed limits. However, the exclusion power
of the shape analysis almost pushes the Higgs-portal coupling into a regime where the two prescriptions
for the Higgs propagator hardly differ any more.

\subsection{Recasting of the cut-and-count analysis}\label{sec:recast}

In this section we perform a LO Monte Carlo recasting of the cut-and-count 
search in \citere{Sirunyan:2018owy}, \ie we compute 
$\sigma_h(q^2)$ using our Monte Carlo setup and derive $\mu_{\mathcal{H}}^{95\%}(q^2)$
using \refeq{eq:limitXS} as explained in the following.
This allows us to validate our Monte Carlo setup, which is, in particular, needed to 
obtain projections for the HL-LHC and the HE-LHC in Section~\ref{sec:projections}.

Since no electroweak corrections are included, the off-shell Higgs cross 
section $\sigma_h(q^2)$ can be obtained by simulating the corresponding on-shell cross
section in the SM with different values for the Higgs mass.
We generate events with \MGfull~v2.6~\cite{Alwall:2014hca} in the 5-flavor scheme.
The renormalization and factorization scales are set to the $W$ mass, 
$\mu_R = \mu_F = M_W$~\cite{deFlorian:2016spz}. 
We use the {\sc NNPDF}3.0~\cite{Ball:2014uwa} LO PDF set with $\alpha_s(M_Z) = 0.118$ provided 
by {\sc LHAPDF} v6.1.6~\cite{Buckley:2014ana}. 
The events are subsequently showered and hadronized using \PY~v8.235~\cite{Sjostrand:2014zea}. 
Detector simulation is performed with the CMS detector card in \DEL~v3.4.2~\cite{deFavereau:2013fsa}. 
Jets are clustered with \FJ~v3.3.1~\cite{Cacciari:2011ma} using the anti-kT 
algorithm~\cite{Cacciari:2008gp} with $R = 0.4$.
To model the \SI{13}{\TeV} CMS VBF analysis~\cite{Sirunyan:2018owy}, we employ the cuts summarized 
in the corresponding column in Table~\ref{tab:cuts}.

Although targeted to select electroweak VBF Higgs production, a subdominant contribution from gluon-initiated
Higgs production (associated with two jets) contaminates the search. We estimate this contribution 
with~\MGfull\ using the HEFT model and reweight~\cite{Mattelaer:2016gcx} the events
to the 1-loop level~\cite{Hirschi:2015iia} 
using the NLO UFO model obtained with \textsc{NloCT}~\cite{Degrande:2014vpa}.
We find that gluon fusion contributes with roughly \SI{20}{\percent} of the VBF production
channel. However, the scale uncertainties for gluon fusion are large. Comparing with the gluon-fusion
contribution of the more elaborate simulation in Figure~6 of \citere{Sirunyan:2018owy}, we find that 
our simulation overshoots by a factor of two. Hence, the actual gluon-fusion contribution of the \SI{13}{\TeV} analysis
is not expected to be much bigger than \SI{10}{\percent}. Moreover, using larger dijet invariant mass cuts, the gluon-fusion
contribution is expected to be even more suppressed for an analysis at the HL- or HE-LHC. Thus, for simplicity,
in the following the gluon-fusion contribution is not simulated explicitly. However, it is indirectly taken into
account by a rescaling of our cross section as discussed in the following.

The CMS cut-and-count analysis~\citere{Sirunyan:2018owy} quotes 743 nominal signal events with an error of 129
events corresponding to a signal cross section (including acceptance times efficiency) of \SI{20.7}{\femto\barn}. 
Our simulation of VBF production yields a fiducial cross section of \SI{14.2}{\femto\barn} with a scale uncertainty of roughly \SI{25}{\percent}. 
To improve our LO result, we rescale the LO signal cross section $\sigma_{h}(q^2)$ by the
corresponding factor 1.46 to match the on-shell Higgs-boson cross section in~\citere{Sirunyan:2018owy}. With this rescaling,
we predict $\mu_\mathcal{H}$ and compare to Figure~7 of~\citere{Sirunyan:2018owy} as shown in~Figure~\ref{fig:calHlim}. 
Up to $m_\mathcal{H}=\SI{150}{\GeV}$ we find agreement below the percent level. 
Our prediction deviates by \SI{6}{\percent} at $m_\mathcal{H}=\SI{200}{\GeV}$ and by \SI{14}{\percent} at \SI{300}{\GeV}. However, close to the resonance,
$m_S\lesssim63\,$GeV, where the 13~TeV analysis is sensitive, the relevant $m_\mathcal{H}$ range
is confined to be well below \SI{200}{\GeV}. For DM masses around \SI{70}{\GeV} still more than \SI{75}{\percent} of the cross section arises from contributions with $m_\mathcal{H}<\SI{200}{\GeV}$.
Hence, good agreement of our 
Monte Carlo study with the results in Figure~\ref{fig:limits13TeV} can be expected. We indeed find
that the results in Figure~\ref{fig:limits13TeV} are reproduced at the level of the statistical 
Monte Carlo error 
not only below but also at and above threshold. 

To validate our analysis by an equivalent, independent calculation, for the cut-and-count analysis we also directly simulate DM production with \MG\ without making use of $\mu_{\mathcal{H}}^{95\%}(q^2)$ and~\refeq{eq:findbound}.
To this end we implement the Higgs-portal model in \FR~v2.3~\cite{Christensen:2008py,Alloul:2013bka} 
and export it to the UFO format~\cite{Degrande:2011ua}. We use the same signal rescaling factor as above.
Note that \MG\ only supports fixed-width propagators.
We use two different ways to overcome this limitation. On the one hand, we simulate the signal using a fixed width and modify 
the weight of each event by the respective ratio of the propagators squared.
On the other hand, we manipulate the $hSS$ vertex in the UFO model to include the propagator ratio.\footnote{Both 
	approaches only work if higher-order corrections in $\lambda_\text{HP}$ are neglected, as it is 
	done everywhere in this work.}
As to be expected, both approaches give consistent results and agree with the results based on \refeq{eq:findbound} 
within MC errors ($\sim$\SI{1}{\percent}) for the event yields.

\begin{table}
	\begin{center}\begin{tabular}{c|c|c}
		$\sqrt{s}$ & \SI{13}{\TeV} & \SI{14}{\TeV} / \SI{27}{\TeV}  \\ \hline
		$p_T^{j_1}$ & $>\SI{80}{\GeV}$ & $>\SI{80}{\GeV}$ \\
		$p_T^{j_2}$ & $>\SI{40}{\GeV}$ & $>\SI{40}{\GeV}$ \\
		$|\eta_j|$ & $<4.7$ & $<5.0$ \\
		$\min\left(|\eta_{j_1}|,|\eta_{j_2}|\right)$ & $<3.0$ & -- \\
		$M_{jj}$ & $>\SI{1.3}{\TeV}$ & $>\SI{2.5}{\TeV}$ / $>\SI{6}{\TeV}$  \\
		$\eta_{j_1}\!\cdot\eta_{j_2}$ & $<0$ & $<0$ \\
		$|\Delta\eta_{jj}|$ & $>4.0$ & $>4.0$ \\
		$|\Delta\phi_{jj}|$ & $<1.5$ & $<1.8$ \\
		$\MET$ & $>\SI{250}{\GeV}$ & $>\SI{190}{\GeV}$ \\
		$|\Delta\phi_{j\MET}|$ & $>0.5\ (p_T^j>\SI{30}{\GeV})$ & $>0.5\ (p_T^j>\SI{30}{\GeV})$ \\
		photon veto & $p_T^\gamma > \SI{15}{\GeV},\ |\eta_\gamma| < 2.5$ & -- \\
		electron veto & $p_T^e > \SI{10}{\GeV},\ |\eta_e| < 2.5$ & $p_T^e > \SI{10}{\GeV},\ |\eta_e| < 2.8$ \\
		muon veto & $p_T^\mu > \SI{10}{\GeV},\ |\eta_\mu| < 2.4$ & $p_T^\mu > \SI{10}{\GeV},\ |\eta_\mu| < 2.8$  \\
		$\tau$-lepton veto & $p_T^\tau > \SI{18}{\GeV},\ |\eta_\tau| < 2.3$ & $p_T^\tau > \SI{20}{\GeV},\ |\eta_\tau| < 3.0$ \\
		$b$-jet veto & $p_T^b > \SI{20}{\GeV},\ |\eta_b| < 2.4$ & $p_T^b > \SI{30}{\GeV},\ |\eta_b| < 5.0$ 
	\end{tabular}\end{center}
	\caption{Analysis cuts used in this paper. The cuts for \SI{13}{\TeV} and \SI{14}{\TeV} are taken 
	from~\cite{Sirunyan:2018owy} and~\cite{CMS:2018tip}, respectively. The cuts for
	the \SI{27}{\TeV} HE-LHC are identical to the ones for the \SI{14}{\TeV} HL-LHC except for the
	cut on $M_{jj}$.
	\label{tab:cuts}}
\end{table}

\section{HL-LHC and HE-LHC projections}\label{sec:projections}

Our HL-LHC projections are based on~\citere{CMS:2018tip}. For an integrated luminosity of
\SI{3}{\per\atto\barn} at \SI{14}{\TeV} center-of-mass energy, the CMS study 
defines an optimal cut-and-count search by the fiducial phase-space 
region as shown in Table~\ref{tab:cuts}. While the cut on the missing energy
has been lowered to \SI{190}{\GeV} compared to the \SI{13}{\TeV} analysis~\cite{Sirunyan:2018owy},
the higher luminosity allows for raising the dijet invariant mass cut to 
\SI{2.5}{\TeV} in order to increase the signal-to-background ratio while still controlling
the background well by data-driven methods.

As in Section~\ref{sec:reinterp}, we make use of the experimental projection 
in~\citere{CMS:2018tip} as much as possible, \ie we employ the cross section 
and the corresponding event number for all the backgrounds. Moreover, in analogy to 
Section~\ref{sec:recast}, we use the
simulated on-shell Higgs-boson production cross section of~\citere{CMS:2018tip}, which is 
based on more sophisticated Monte Carlo simulations, to rescale our simulation based on LO Monte Carlo and \DEL.
For the \SI{14}{\TeV} case, the corresponding rescaling factor is given by 1.54.
If we had used the rescaling factor found for the \SI{13}{\TeV} analysis in Section~\ref{sec:recast},  
our prediction for the SM Higgs-production cross section would have deviated from the cross section in \citere{CMS:2018tip} by less than \SI{5}{\percent}.
Hence, the rescaling
factor does not vary substantially with energy and the details of the analysis.  
We assume that this is also the case for the HE-LHC analysis and will later use the \SI{14}{\TeV} rescaling 
factor also at \SI{27}{\TeV}.

\begin{figure}[t]
\centering
\includegraphics[bb=0 0 225 155]{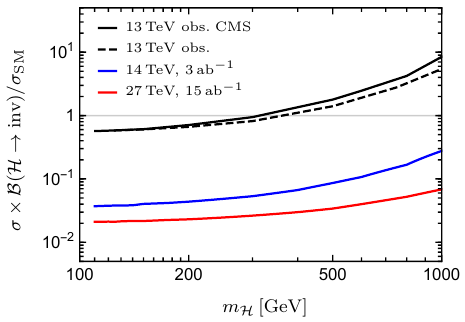}
\caption{\SI{95}{\percent}~CL upper limit on $\mu_\mathcal{H}= \sigma\times {\cal B} ({\cal H}\to\text{inv}) /\sigma_\text{SM}$
for an additional SM-like Higgs boson ${\cal H}$ that does not mix with the SM Higgs boson as a function of its mass, $m_{\cal H}$. 
The black solid curve shows the existing limit from CMS reported in~\citere{Sirunyan:2018owy} while the black dashed curve shows our limit from the recasting of this search. The blue and red curves show our projected limits for the \SI{14}{\TeV} HL-LHC and
\SI{27}{\TeV} HE-LHC, respectively. 
}
\label{fig:calHlim}
 \end{figure}

\begin{figure}[t]
\centering
\includegraphics[bb=0 0 265 195]{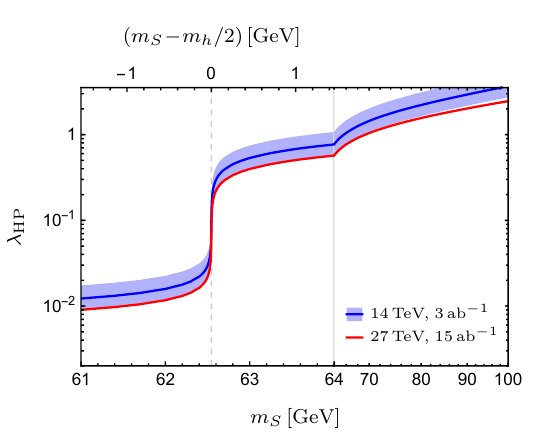}
\caption{%
Projected \SI{95}{\percent}~CL upper limits on the Higgs-portal coupling for the \SI{14}{\TeV} HL-LHC (blue curve) and \SI{27}{\TeV} HE-LHC (red curve) assuming an integrated
luminosity of \SI{3}{\per\atto\barn} and \SI{15}{\per\atto\barn}, respectively. The shaded band around the blue curve indicates the sensitivity of the respective limit to the systematic error on the background from~\citere{CMS:2018tip} which is multiplied by a factor of 2 (1/2) for the upper (lower) boundary of the band. 
To resolve the resonant region, for the abscissa we impose a scale break at a DM mass of \SI{64}{\GeV}, below (above) which we plot the mass linearly (logarithmically).
\label{fig:projectedlimits}
}
\end{figure}

To obtain our HL-LHC projections we
calculate $\sigma_h(q^2)$ with our rescaled LO Monte Carlo simulation and derive the 
corresponding limit on the signal strength $\mu_\mathcal{H}$ as a function of the additional 
Higgs mass $m_\mathcal{H}$. The limit $\mu_\mathcal{H}$ is shown as the blue curve in Figure~\ref{fig:calHlim}. 
In analogy to Section~\ref{sec:reinterp}, we derive constraints on the Higgs-portal coupling 
using \refeq{eq:findbound}. 
The resulting projected limits on the Higgs-portal coupling are shown in Figure~\ref{fig:projectedlimits} as the blue curve. 
They exclude $\lambda_\text{HP}\ge 0.09$ (0.9) for $m_S=m_h/2$ (\SI{65}{\GeV}) at \SI{95}{\percent}~CL\@.
As the analysis relies on a data-driven background prediction, its relative systematic uncertainty, $\sigmaSYS$, is
expected to be small. From~\citere{CMS:2018tip} we derive $\sigmaSYS=\SI{1.4}{\percent}$ (see below for more details).
We indicate the dependence on the systematic uncertainties of the background prediction by the blue shaded band,
for which we vary $\sigmaSYS$ by a factor of 2.
Note that the expected limit of the high-luminosity cut-and-count analysis supersedes the observed limits from the 
\SI{13}{\TeV} shape analysis already for a systematic uncertainty below \SI{12}{\percent}.
Of course, a more sophisticated (shape-like) 
analysis is expected to even further improve the limits as for the \SI{13}{\TeV} search.

\begin{figure}[t]
\centering
\includegraphics[bb=0 0 440 160]{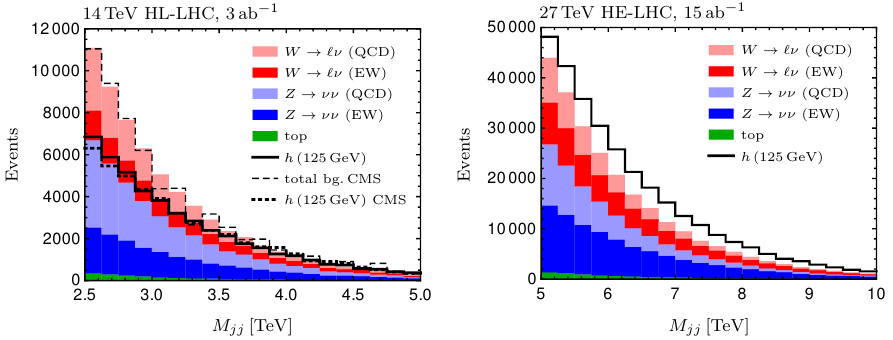}
\caption{%
Binned $M_{jj}$ distributions for our background (stacked histograms) and signal (thick solid line) prediction
for the \SI{14}{\TeV} HL-LHC (left panel) and \SI{27}{\TeV} HE-LHC (right panel). Apart from $M_{jj}$, the analysis cuts listed in Table~\ref{tab:cuts} are applied. For comparison, we show the total background (thin dashed line) and signal (thick dotted line) predictions for the \SI{14}{\TeV} HL-LHC from CMS as reported in~\citere{CMS:2018tip}.
\label{fig:mjjdist}  
}
\end{figure}

To evaluate the potential of the HE-LHC, we now turn to projections for a \SI{27}{\TeV} machine.
Since there is no corresponding experimental projection for a \SI{27}{\TeV} HE-LHC option, we  
simulate the dominating backgrounds and estimate the respective error which is most crucial 
to assess the sensitivity of the search.
To this end we exploit the corresponding information provided in the \SI{14}{\TeV} HL-LHC projection~\cite{CMS:2018tip}
as explained in the following.
We generate the leading background processes $p p \to Z + \text{jets}$ and $p p \to W^\pm + \text{jets}$ for \SI{14}{\TeV}
at LO using \MG\ and employ the MLM matching scheme~\cite{alwall-2008-53} to merge samples with 2 and 3 jets.
We separately simulate processes with jets produced through electroweak~(EW) interactions and processes in which all jets
originate from QCD radiation, neglecting interference between the two.
The detector response is simulated using the HL-LHC detector card in \DEL. 
To improve our LO background calculation in analogy with our signal prediction, we determine
rescaling factors for each background process by comparing the cross sections (including acceptance times efficiency) to the 
cross sections reported in~\citere{CMS:2018tip}. 
By rescaling to the results in~\citere{CMS:2018tip}, we profit both
from the more elaborate background simulation and the more sophisticated simulation of detector effects
in the experimental projection. We include the subleading top-contribution (below \SI{4}{\percent}) 
in the rescaling factor for the $W$ backgrounds assuming a roughly similar $M_{jj}$ dependence. 
We then simulate the various backgrounds at \SI{27}{\TeV} and perform
a rescaling for each channel with the rescaling factors determined at \SI{14}{\TeV}. 
The corresponding $M_{jj}$ distributions are shown in Figure~\ref{fig:mjjdist}.

Given the large statistics, the systematic uncertainty for the background prediction is crucial for the 
sensitivity of the corresponding search. The background prediction at a \SI{27}{\TeV} HE-LHC in the experimental search
will be data driven using control regions as it is the case for the \SI{13}{\TeV} search discussed in Section~\ref{sec:13TeVlim} and 
in the \SI{14}{\TeV} HL-LHC projection. For example, the background due to $Z$-boson production, where the $Z$ bosons decay into
neutrinos, can be measured to a large extent by $Z$-boson decays into charged leptons in a control 
region. 
Due to the smaller branching ratio into charged leptons, this measurement has smaller statistics.
The corresponding statistical error 
scales like the square root of the number of events in the control region.
In addition, there is always a relative systematic error 
that does not scale with the integrated luminosity and arises from relating the control-region measurements to the signal-region
background estimate. We denote this luminosity-independent error by $\sigmaIND$.
While a full account of this procedure is clearly beyond the scope of this work and can only be
performed in the experimental analysis, the above discussion motivates the following simple modeling for the
relative systematic error $\sigmaSYS$ of the background prediction according to
\begin{equation}
\sigmaSYS = \sqrt{ f/N_\text{B} + (\sigmaIND)^2}\,,
\end{equation}
where $N_\text{B}$ is the number of background events in the signal region and the parameter $f$ reflects the fact that 
the control-region measurement potentially has smaller statistics (in which case $f>1$)
but still scales like a statistical error with increasing luminosity. 
Hence, the \SI{95}{\percent}~CL expected limit for the number of signal events $N_\text{S}$ can be obtained in the asymptotic (Gaussian) limit from
\begin{equation}
\frac{N_\text{S}}{\sqrt{N_\text{B} + \left(\sigmaSYS N_\text{B}\right)^2  + N_\text{S}}} = 1.96 \, .
\end{equation}
To obtain realistic estimates for $f$ and $\sigmaIND$, we use the \SI{14}{\TeV} projection.
As $N_\text{S}$ and $N_\text{B}$ scale with the integrated luminosity we use the limit for the invisible 
branching ratio for the three
different luminosities provided in Figure 5 of~\citere{CMS:2018tip} to determine 
$f=1.5$ and $\sigmaIND=\SI{1.3}{\percent}$.
Using these values, our simple modeling of
the systematic error nicely reproduces the results in Figure~5 of~\citere{CMS:2018tip}.
We use these numbers as the best estimate parametrising
the background uncertainty for the HE-LHC\ projection. 

\begin{figure}[t]
\centering
\includegraphics[bb=0 0 230 150]{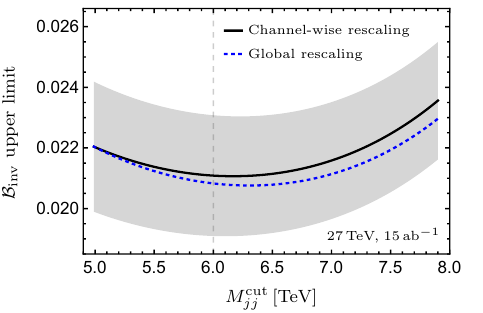}
\caption{Upper limit on the invisible branching ratio of the SM Higgs boson
as a function of the $M_{jj}$-cut for the \SI{27}{\TeV} HE-LHC\@. Otherwise cuts from Tab.~\ref{tab:cuts} are applied.
We show the limit using a channel-wise (black, solid) and global (blue, dotted) rescaling of the background prediction,
see text for details. 
For illustration, the grey band shows the respective 
shift of the solid black line resulting from a $\SI{10}{\percent}$ variation of the background cross section. 
\label{fig:mjjdep}
}
\end{figure}

Optimizing the cut-and-count search for a \SI{27}{\TeV} machine, we find the biggest potential for gaining sensitivity 
in strengthening the cut on $M_{jj}$, while
we leave all other cuts as in the \SI{14}{\TeV} analysis.\footnote{%
A further improvement might be achieved by increasing the cut on $|\Delta\eta_{jj}|$. However, the potential gain sensitively 
depends upon the detector performance in the region of large $\eta$.
}
In order to obtain an estimate for the optimal 
$M_{jj}$-cut at \SI{27}{\TeV} we employ the following strategy. 
Producing a high-statistics sample with a detector-level $M_{jj}$-cut at \SI{5}{\TeV} we find the 
$M_{jj}$ distribution shown in the right panel of Figure~\ref{fig:mjjdist}, which allows us to
compute the total number of background events as a function of a given $M_{jj}$-cut.
Furthermore, we simulate a high-statistics sample for on-shell SM Higgs production
assuming \SI{100}{\percent} invisible branching ratio. The resulting branching ratio limit as a function of $M_{jj}$ is shown in Figure~\ref{fig:mjjdep}. 
For comparison, we also show the result using a global 
rescaling factor, \ie rescaling all simulated background contributions with the same factor instead of rescaling each background channel individually.
This global rescaling factor is
chosen such that the two background estimates coincide for a $M_{jj}$-cut at \SI{5}{\TeV}. The small difference between a global and channel-wise rescaling is an indicator for the robustness of the $M_{jj}$ dependence of our background estimate.
We find that the upper limit on the invisible branching ratio is strongest at $M_{jj}$-cuts around \SIrange{6}{6.5}{\TeV}, with little sensitivity on the detailed choice. We adopt the 
value $M_{jj}>\SI{6}{\TeV}$. For this choice, the relative uncertainty of the background $\sigmaSYS=\SI{1.4}{\percent}$ for an integrated luminosity of \SI{15}{\per\atto\barn} 
is already dominated by $\sigmaIND$.  

Note that the uncertainty of our LO Monte Carlo simulation in the prediction of the background yield 
is large, around $10\!-\!\SI{20}{\percent}$, which translates into a shift in the limits by roughly the same amount (see Figure~\ref{fig:mjjdep}). 
However, this uncertainty disappears, of course, once data has been taken, \ie the data-driven background prediction is made.

Using the above input on the background prediction and the corresponding error estimate, we calculate
$\sigma_h(q^2)$ with our rescaled LO Monte Carlo simulation as before and derive the 
corresponding limit on the signal strength $\mu_\mathcal{H}$ shown in Figure~\ref{fig:calHlim}.
For the SM Higgs with $m_h\simeq\SI{125}{\GeV}$ the projected HE-LHC \SI{95}{\percent}~CL upper limit on the invisible branching ratio is 0.021.
The resulting limits on the Higgs-portal coupling are shown in Figure~\ref{fig:projectedlimits}.
For $m_S\lesssim\SI{61}{\GeV}$ the Higgs portal coupling is expected to be constrained to less than 
$\lambda_\text{HP}=0.01$, while at the resonance $m_S=m_h/2$ we find $\lambda_\text{HP}<0.077$. Useful (perturbative) limits can be obtained up to DM masses of \SI{110}{\GeV}.
These limits can be viewed as conservative. Improving the systematic uncertainty in the background prediction or
using more sophisticated shape-like or multivariate techniques might further strengthen the limits. 

\section{Conclusion}\label{sec:summary} 

We have conducted a dedicated study of scalar Higgs-portal DM in the VBF channel, presenting limits 
from current LHC data as well as projections for the HL- and HE-LHC upgrades.
Results for other types of Higgs-portal DM candidates are included in
Appendix~\ref{app:othmod}.
Due to its distinct topology, the VBF channel provides particularly promising prospects to probe this kind of models.
The analysis is based on the VBF search for invisible Higgs decays~\cite{Sirunyan:2018owy} and the corresponding 
high-luminosity forecast~\cite{CMS:2018tip} by CMS\@.
Our projected sensitivities include an estimate of the systematic uncertainty 
that can be achieved with a data-driven background prediction.

Special focus has been put on the DM mass region $m_S \simeq m_h/2$ as this region is
particularly well-motivated due to constraints from the DM abundance and direct detection.
If the mass of DM is slightly too large for being produced via an on-shell Higgs boson, the invisible channel 
opens up within the Higgs-boson resonance. This gives rise to an unphysical enhancement of the cross section if the total Higgs-boson width is kept 
fixed in the propagator. 
We solve this problem by using a running-width prescription.
For the observed limits on the Higgs-portal coupling, the fixed-width prescription over-estimates the 
constraining power of the cut-and-count (shape) analysis by up to roughly \SI{30}{\percent} (\SI{15}{\percent}).
Note that an effect of similar size would also be present in the annihilation cross section relevant for 
the computation of the freeze-out abundance, if similarly large couplings were considered.

We obtain a \SI{95}{\percent}~CL upper limit on the Higgs invisible branching ratio of 0.021 at the HE-LHC with \SI{15}{\per\atto\barn}.
The corresponding \SIlist[list-units=single]{13;14}{\TeV} limits are 0.3~\cite{Sirunyan:2018owy} and 0.038~\cite{CMS:2018tip}, respectively.
With current LHC data, Higgs-portal couplings of the order of $\lambda_\text{HP} \simeq 0.04$ 
(at $m_S = \SI{61}{\GeV}$) can be excluded below threshold. 
At $m_S = m_h/2$ the LHC probes $\lambda_\text{HP} \simeq 0.3$, whereas above the threshold only couplings 
as large as $\lambda_\text{HP} \simeq 2.5$ (at $m_S = \SI{64}{\GeV}$) can be reached.
With an integrated luminosity of \SI{3}{\per\atto\barn} collected at the HL-LHC the corresponding limits improve to 
$\lambda_\text{HP} \simeq 0.01,\ 0.09,\ \text{and}\ 0.8$, respectively.
At the HE-LHC with \SI{15}{\per\atto\barn} we estimate a further improvement of these bounds by roughly \SI{30}{\percent}.
Stronger limits may be achieved employing analysis techniques beyond a simple cut-and-count analysis
and/or by improving the systematic uncertainty.

Following \citere{Sirunyan:2018owy}, we also present our HL- and HE-LHC results as upper limits on the signal strength $\mu_\mathcal{H}$ of an invisibly decaying, SM-like Higgs boson with mass $m_\mathcal{H}$.
We find that at the HL-LHC (HE-LHC) $\mu_\mathcal{H}$ can be constrained to values better than $\mu_\mathcal{H} < 0.1$ for masses below $m_\mathcal{H} \lesssim \SI{500}{\GeV}\ (\SI{1}{\TeV})$.
These limits allow for a simple reinterpretation for other Higgs-mediated DM models using \refeq{eq:findbound}.
All results can be found in digital form in the supplementary material to this paper.

\section*{Acknowledgements}

We would like to thank Christian Schwinn, Pedro Schwaller, and Felix~Yu for helpful discussions. 
We acknowledge support by the German Research Foundation DFG through the research unit ``New physics at the LHC'', the CRC/Transregio 257 
``P3H: Particle Physics Phenomenology after the Higgs Discovery'' 
and the Cluster of Excellence ``Precision Physics, Fundamental Interactions, and Structure of Matter'' (PRISMA+ EXC 2118/1) 
within the German Excellence Strategy (Project ID 39083149).
J.H.~acknowledges support from the F.R.S.-FNRS, of which he is a postdoctoral researcher.
E.M.~acknowledges the computing time granted on the supercomputer Mogon at Johannes Gutenberg University Mainz. 

\begin{appendix}

\section{Threshold at a resonance}\label{app:res}
\label{sec:resonance}

It is common to use a fixed-width prescription for the resonant Higgs propagator, \ie
\begin{equation}
P_\text{f}(q^2) = \frac{\I}{q^2-m_h^2+\I\, m_h \,\Gamma_\tot(m_h^2)} \, ,
\end{equation}
where $m_h$ is the Higgs-boson mass and $\Gamma_\tot(m_h^2)$ is the total width of an on-shell Higgs 
boson. The production rate of all Higgs-boson decay modes via a resonant Higgs boson is then 
given by
\begin{equation}
\label{eq:fixedwidthcrosssection}
\sigma^\tot_\text{f} = \int \frac{\D q^2}{2\pi} \, \sigma_{h}(q^2) \, 
\frac{2q\Gamma_\tot(q^2)}{\left(q^2-m_h^2\right)^2 + m_h^2\Gamma_\tot^2(m_h^2)} \,,
\end{equation}
where $\sigma_{h}(q^2)$ is the production cross-section for a given Higgs-boson invariant 
mass $q^2$. The integral is dominated by the on-shell region $q^2\sim m_h^2$ which has a width of
$\mathcal{O}(m_h\Gamma_\tot)$. Hence, if $\Gamma_\tot(m_h^2)$ is small and $\sigma_h(q^2)$
as well as $\Gamma_\tot(q^2)$ are smooth functions, the narrow-width approximation 
\begin{equation}
\label{eq:fixedw}
\sigma^\tot_\text{f} \simeq \int \frac{\D q^2}{2\pi} \, \sigma_h(m_h^2) \, 
\frac{2m_h\Gamma_\tot(m_h^2)}{\left(q^2-m_h^2\right)^2 + m_h^2\Gamma_\tot^2(m_h^2)} = 
\sigma_h(m_h^2)
\end{equation}
is valid. However, if the total width $\Gamma_\tot(q^2)$ is a rapidly varying
function in the resonance region, the fixed width prescription for the propagator and the 
narrow-width approximation may break down. In particular, if a new decay channel with a large coupling 
opens up close to the resonance, the increase in $\Gamma_\tot(q^2)$ leads to a large increase of 
$\sigma^\tot_\text{f}$ if it is calculated using \refeq{eq:fixedwidthcrosssection}. 
Hence, $\sigma^\tot_\text{f}$ is not
related any more to the production cross section $\sigma_h(m_h^2)$ as it should be. 

\begin{figure}[!ht]
\centering
\includegraphics[bb=0 0 230 170]{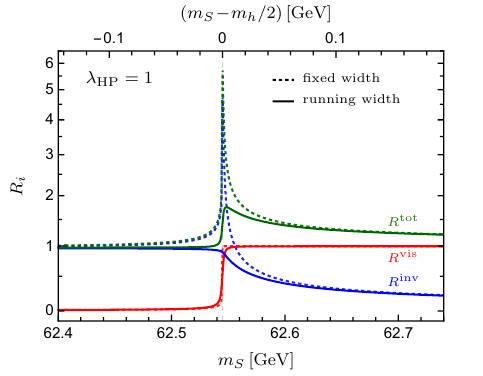}
\caption{Cross section ratios $R_\text{f}\approx \frac{\sigma_\text{f}}{\sigma_h(m_h^2)}$ (dashed lines)
and $R_\text{r}\approx \frac{\sigma_\text{r}}{\sigma_h(m_h^2)}$ (solid lines) using a fixed- and a
running-width prescription for the Higgs-boson propagator, respectively. We consider the ratios $R^\tot$ w.r.t.\ the total 
cross section (black), $R^\mathrm{vis}$ w.r.t.\ the SM Higgs-decay products (red) and 
$R^\mathrm{inv}$ w.r.t.\ the Higgs decay into singlets (blue). This plot is obtained for 
$\lambda_\text{HP}=1$.}
\label{fig:B_m_S}
\end{figure}

To illustrate the issue, one can use the narrow-width approximation for the production cross 
section and investigate the ratio 
\begin{equation}
R_\text{f}^\tot=
\int \frac{\D q^2}{2\pi} \, 
\frac{2 q \Gamma_\tot(q^2)}{\left(q^2-m_h^2\right)^2 + m_h^2\Gamma_\tot^2(m_h^2)} 
\simeq \frac{\sigma^\tot_\text{f}}{\sigma_h(m_h^2)} \,.
\end{equation}
In the following, we use the tree-level width for the Higgs decay into two singlets
\begin{equation}
\Gamma_\text{inv}(q^2) = 
\frac{\lambda_\text{HP}^2}{32\pi}\frac{v^2}{q}\sqrt{1-\frac{4 m_S^2}{q^2}} \, \Theta(q^2-4 m_S^2)\,
\end{equation}
to define the total width $\Gamma_\tot(q^2)=\Gamma_\mathrm{vis} + \Gamma_{\text{inv}}(q^2)$, where 
$\Gamma_\mathrm{vis}$ is the Higgs width in the SM\@.
We neglect the $q^2$ dependence of $\Gamma_\mathrm{vis}$ as the dominant effect comes from the invisible part.
Figure~\ref{fig:B_m_S} shows the ratio $R_f^\tot$ as a function of the singlet
mass $m_S$. If the invisible decay channel into singlets opens up in the vicinity of the Higgs-boson
resonance, the description is clearly unphysical since $R_\text{f}^\tot$ can become large. 

The obvious improvement is to use a running-width prescription in the Higgs propagator according to 
\begin{equation}
P_\text{r}(q^2) = \frac{\I}{q^2-m_S^2+\I \sqrt{q^2} \,\Gamma_\tot(q^2)} \, .
\end{equation}
The ratio
\begin{equation}
R_\text{r}^\tot=
\int \frac{\D q^2}{2\pi} \, 
\frac{2 q \Gamma_\tot(q^2)}{\left(q^2-m_h^2\right)^2 + q^2\Gamma_\tot^2(q^2)} 
\approx \frac{\sigma^\tot_\text{r}}{\sigma_h(m_h^2)}
\end{equation}
is well behaved as can be seen in Figure~\ref{fig:B_m_S}. In particular, if the total cross section
is written as the sum of the production cross-section of all SM final-states $\sigma^\mathrm{vis}$ 
and the singlet final-state $\sigma^\mathrm{inv}$ the physics interpretation of the results becomes
transparent. If the decay channel to singlets is already open at the Higgs-boson resonance, it dominates
for large coupling and almost all the produced Higgs bosons decay into the invisible channel. If,
on the other hand, the decay channel to singlets is not open, 
$\sigma^\mathrm{vis}$ is almost
unchanged with respect to the SM and there is additional off-shell production of the singlet
final-state. In the resonance region, the running width in the Higgs propagator leads to a smooth
transition between the on-shell and the off-shell production of the singlet final-state 
$\sigma^\mathrm{inv}$ as shown in Figure~\ref{fig:B_m_S}. 

\section{Fermion, vector and tensor dark matter}\label{app:othmod}

In this appendix we derive \SI{95}{\percent}~CL upper limits on the Higgs portal coupling for other
choices of DM particles. We
consider the interaction Lagrangians (before electroweak symmetry breaking) 
\begin{subequations}
	\label{eq:HPM-effLags}
  \begin{equation}
    \Lag_\text{HP} = 
      - \frac{\lamchi}{\Lambda} \Phi^\dagger \Phi \overline{\chi} 
      \chi
      \label{eq:HPM-FermionLag}
  \end{equation}
for a Majorana fermion $\chi$\footnote{%
Note that we do not consider a possible pseudo-scalar coupling or a coupling to the hypercharge field-strength tensor
that are also allowed at dimension 5.},
  \begin{equation}
    \Lag_\text{HP} = %
     - \frac{\lamX}{2} \,\Phi^\dagger\Phi\,X^\mu X_\mu 
    \label{eq:HPM-effVectorLag}
  \end{equation}
for a vector $X$, and 
   \begin{equation}
    \begin{aligned}
      \Lag_\text{HP} =
        - \frac{\lamB}{2}\,\Phi^\dagger \Phi\,B^{\mu\nu} B_{\mu\nu}
    \end{aligned}
    \label{eq:HPM-effTensorLag}
  \end{equation}
\end{subequations}
for an anti-symmetric rank-2 tensor $B$. 
More details about the models can be found in~\cite{Kanemura:2010sh,Endo:2014cca,Cata:2014sta}. For the tensor model, we
follow the conventions of \citere{Cata:2014sta}.
The Higgs-portal interaction leads to the following expressions for the invisible Higgs-boson width~\cite{Kanemura:2010sh,Endo:2014cca}:
  \begin{subequations}\begin{align}
  \Gamma\left(h\rightarrow \chi \overline{\chi}\right)
  =&\ \frac{\lamchi^2 v^2}{4\pi\Lambda^2} m_{h} \left(1-\frac{4 m_\chi^2}{m_{h}^2}\right)^{3/2}  ,\\
  \Gamma\left(h\rightarrow X X\right)
  =&\ \frac{\lamX^2}{128\pi} \frac{v^2}{m_{h}} \frac{m_{h}^4 - 4 m_{h}^2 m_X^2 + 12 m_X^4}{m_X^4} \sqrt{1-\frac{4 m_X^2}{m_{h}^2}} ,\\
  \Gamma\left(h\rightarrow B B\right)
  =&\ \frac{\lamB^2}{16\pi} \frac{v^2}{m_{h}} \frac{m_{h}^4 - 4 m_{h}^2 m_B^2 + 6 m_B^4}{m_B^4} \sqrt{1-\frac{4 m_B^2}{m_{h}^2}}.
  \end{align}\end{subequations}

\begin{figure}[t]
\centering
\includegraphics[bb=0 0 450 330]{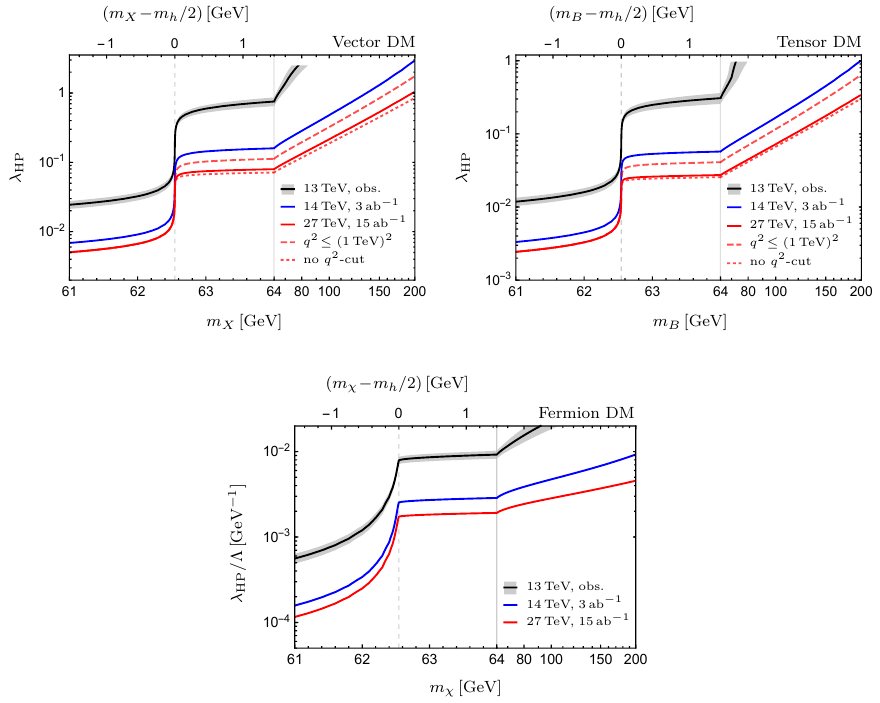}
\caption{%
\SI{95}{\percent}~CL upper limits on the Higgs-portal coupling for the case of vector (upper left), tensor (upper right) and Majorana fermion
DM (lower panel). The black curve (and grey band) denotes the \SI{13}{\TeV} LHC observed limit (and the signal uncertainty). 
The blue and red curves denote the \SI{14}{\TeV} HL- and \SI{27}{\TeV} HE-LHC projections, respectively. For the vector and tensor models, we use
a cut-off for the integral in \refeq{eq:findbound} as discussed in the text.
The dashed and dotted light red curves indicate the HE-LHC limits if a \SI{1}{\TeV} cut-off or no
cut-off is used instead, respectively.
To resolve the resonant region, for the abscissa we impose a scale break at a DM mass of \SI{64}{\GeV}, below (above) which we plot the mass 
linearly (logarithmically).
\label{fig:limitsXBchi}
}
\end{figure}

In contrast to the case of singlet scalar DM treated in the main text, the models in~\refeq{eq:HPM-effLags} considered in this appendix are not UV complete.
They are non-renormalizable and violate perturbative unitarity at high energies.
As a consequence, the DM production cross section \refeq{eq:DMproduction} receives large contributions from $q^2 \gg m_\text{DM}^2$. 
For example, it is well known from vector-boson scattering in the SM 
that perturbative unitarity is violated at 
$q^2 \sim (\SI{1}{\TeV})^2$ for $m_X\sim m_Z$ and a coupling $\lamX$ of weak size~\cite{Lee:1977eg}.
The unitarity violating contributions are, however, expected to be suppressed in UV-completions of the models, \eg via 
additional degrees of freedom that unitarize the theory.
To exclude the potentially unitarity-violating high-energy contribution, we derive the unitarity limit from $hh\to XX$ scattering at 
$\mathcal{O}(\lamX)$~\cite{Lebedev:2011iq} and cut off the integral in \refeq{eq:DMproduction} at 
$q^2=32 \pi m_X^2/\lamX$ for the vector case. In analogy, we use $q^2=16 \pi m_B^2/\lamB$ in the tensor case.
Note that in the vector and tensor model for $m_{X,B}>m_h/2$ the limits depend approximately linearly on the choice of the cut-off near $q^2 = (\SI{1}{\TeV})^2$. 
Figure~\ref{fig:limitsXBchi} therefore also shows the HE-LHC limits obtained
for a cut-off at $\SI{1}{\TeV}$ (dashed) as a conservative limit and without cut-off (dotted).
For the fermion case, a strong cut-off dependence is absent and we do not employ a cut-off.
Note that we use the running-width prescription which has a unitarizing effect (see also~\citere{Khoze:2017tjt}) if the total width becomes so large  that it dominates the denominator of the 
Higgs-boson propagator. This is also the reason why 
the \SI{13}{\TeV} exclusion lines in the vector and tensor case stop around $m_{X/B} = \SI{75}{\GeV}$. 
The cross-section reaches a maximum for a certain value of $\lambda_\text{HP}$ and decreases again for larger couplings.
To establish limits in this region of parameter space, the high-energy behaviour of the models needs to be studied in more detail which is beyond
the scope of this work.

Using the method of Sec.~\ref{sec:reinterp} 
we compute the limit on $\lambda_\text{HP}$ within the
three models. Figure~\ref{fig:limitsXBchi} shows the respective observed \SI{13}{\TeV} LHC limit from the 
shape analysis (black curves) as well as the projections for the \SI{14}{\TeV} HL-LHC and \SI{27}{\TeV} HE-LHC, respectively\@.
The grey band around the \SI{13}{\TeV} limit denotes the signal uncertainty, see Sec.~\ref{sec:reinterp}.
In the vector model, current data excludes couplings around $\lamX \simeq 0.03\ (0.7)$ slightly below (above) the threshold,
whereas the corresponding limits in the tensor model are smaller by about a factor of 2.
At $m_\text{DM} = m_h/2$, $\lambda_\text{HP} = 0.17\ (0.08)$ is excluded in the vector (tensor) case.
For fermionic DM, $\lamchi/\Lambda \simeq \SIrange[range-units=brackets,range-phrase=\!-\!]{0.5}{10}{\per\TeV}$ ($\lamchi/\Lambda=\SI{8.0}{\per\TeV}$) can be probed for $m_\chi \lesssim \SI{65}{\GeV}\ (m_\chi=m_h/2)$.
The HL- and HE-LHC improve these bounds by a factor of 3 to 10.

\end{appendix}

\addcontentsline{toc}{section}{References}
\bibliography{biblio}
\bibliographystyle{JHEP}

\end{document}